\documentclass[12pt]{article}

\usepackage[dvips]{graphicx}

\newcommand{\nn}{\nonumber}
\newcommand{\be}{\begin{equation}}
\newcommand{\ee}{\end{equation}}
\newcommand{\ba}{\begin{eqnarray}}
\newcommand{\ea}{\end{eqnarray}}
\newcommand{\ci}[1]{\cite{#1}}
\def\={\,=\,}

\newcommand{\LQCD}{\Lambda_{\rm{QCD}}}
\def\als{\alpha_s}

\def\mev{\,{\rm MeV}}
\def\gev{\,{\rm GeV}}

\newcommand{\da}{{distribution amplitude}}
\newcommand{\wf}{wave function}
\newcommand{\lsim}{\raisebox{-4pt}{$\,\stackrel{\textstyle
                                                         <}{\sim}\,$}}

\def\taub{\bar{\tau}}
\newcommand{\tw}{\textwidth}                          
\newcommand{\req}[1]{(\ref{#1})}

\def\sh{\hat{s}}
\def\uh{\hat{u}}
\def\th{\hat{t}}

\begin{document}
\thispagestyle{empty}
\begin{flushright}
RBI-ThPhys-2021-39\\
WUB/21-2\\
November, 15 2021\\[20mm]
\end{flushright}

\begin{center}
{\Large\bf Wide-angle photoproduction of the $\eta'$-meson and 
        its gluon content}\\[0.3em]
\vskip 15mm
P.\ Kroll \\[1em]
{\small {\it Fachbereich Physik, Universit\"at Wuppertal, D-42097 Wuppertal,
Germany}}\\[1em]
K.\  Passek-Kumeri\v{c}ki  \\[1em]
{\small {\it Division of Theoretical Physics, Rudjer Bo\v{s}kovi\'{c} Institute, 
HR-10002 Zagreb, Croatia}}\\

\end{center}

\begin{abstract}
  We investigate wide-angle photoproduction of the $\eta'$ meson within the handbag approach
  to twist-3 accuracy. It turns out that, due to the gluon content of the $\eta'$, this
  process is dominated by twist 2 in contrast with pion and $\eta$ photoproduction. Using
  the presently available information on the twist-2 and twist-3 \da s and on the
  $\eta - \eta'$ mixing, we provide prediction for the $\eta'$ cross section and helicity
  correlations. It is argued that $\eta'$ photoproduction is well suited to improve our
  knowledge of the two-gluon \da{}.
\end{abstract}  
\section{Introduction}
The $\eta'$-meson is a complicated object. It is mainly a flavor-singlet state with
a small admixture of a flavor-octet component. This fact leads to the familiar
$\eta - \eta'$ mixing. A further complication of the description of the $\eta'$-meson
is that, at the twist-2 level, there are two Fock components contributing to the
flavor-singlet state, the quark-antiquark one and the two-gluon one. The associated
\da s, $\Phi_{q1}(\tau,\mu_F)$ and $\Phi_g(\tau,\mu_F)$ mix under evolution.
Here, $\tau$ is a momentum fraction and $\mu_F$ denotes the factorization scale.
Moreover, it is expected that the twist-3 contribution, 
which includes 2- and 3-body ($q\bar{q}g$) Fock components, 
also plays an important role since this is the case for pion and eta
photoproduction \ci{KPK18} as well as deeply virtual electroproduction 
\ci{GK6,bedlinskiy,defurne}.

An accurate determination of the two-gluon \da{} is of utmost importance since the
corresponding Fock component of the $\eta'$, and to a lesser extent of the $\eta$,
plays a role in many hard processes involving these mesons. Thus, the $g^*g^*\eta' (\eta)$
vertex substantially contributes to decay processes such as $\Upsilon(1S)\to \eta'X$
\ci{ali03} . It also contributes to the inclusive \ci{cisek21} and exclusive central
\ci{harland11}  production of the $\eta'$ in high-energy proton-proton collisions at the LHC.
The $B\to \eta'$ form factor, appearing in $B$-meson decays into channels involving the
$\eta'$, is affected by the $gg$ Fock component of the $\eta'$-meson too \ci{ball07}. More
information on the role of that Fock component can be found in the review by Bass and Moskal
\ci{bass18}.

Information on the two-gluon \da{} can be extracted from the $\eta' (\eta)$- photon transition
form factor in a leading-twist analysis to next-to-leading order (NLO) of QCD \ci{KPK03,KPK13}.
However, the present data on these form factors \ci{CLEO,L3,Babar} allow only
to determine the first Gegenbauer coefficients, $a_2^i$ and $a_2^g$, of the quark and gluon
\da s ($i=1,8$)~\footnote{
  Particle independence of the \da s is assumed as in \ci{KPK03,KPK13}. Since in hard
  processes only small spatial quark-antiquark (gluon-gluon) separations are relevant 
  it seems plausible to embed  the particle dependence solely in the decay constants, see
  also \ci{FKS1}.}~\footnote{
    In order to facilitate comparison with other work we have changed 
the definition of the gluon
            \da{} \ci{KPK19} compared to our previous work \ci{KPK03,KPK13}.}   
\ba
\Phi_{qi}(\tau,\mu_F)&=& 6\tau(1-\tau)\,\Big[1 + \sum_{n=2,4,\ldots} a_n^i(\mu_F) C_n^{3/2}(2\tau-1)\Big]\,,\nn\\
\Phi_g (\tau,\mu_F)&=& 30\tau^2(1-\tau)^2\,\sum_{n=2,4,\ldots} a_n^g(\mu_F) C_{n-1}^{5/2}(2\tau-1)\,.
\label{eq:das}
\ea
The Gegenbauer coefficients depend on the factorization scale, $\mu_F$, and the flavor-singlet coefficients mix
with the gluon ones under evolution \ci{grozin}. \\
The Gegenbauer coefficients $a_2$ obtained in \ci{KPK13}
and evolved to the scale $~{\mu_0=2\,\gev}$, 
take the following values 
\ba
a_2^8(\mu_0)&=&-0.039\pm 0.016\,, \nn\\
a_2^1(\mu_0)&=&-0.057\pm 0.012\,, \qquad a_2^g(\mu_0)\=0.38\pm 0.10
\,.
\label{eq:a2}
\ea
The contribution $\propto a_2^g$ is small for the $\gamma \eta'$ form factor since
it is suppressed by the strong coupling, $\als$.
The coefficients \req{eq:a2} are to be regarded as effective ones; they may be contaminated by contributions
from higher order coefficients. Data on the $\gamma^*\eta'(\eta)$ form factor would, in principle,
allow for an extraction of 
the first few real
Gegenbauer coefficient since, for this form factor, the order $n$
coefficients are suppressed by $\omega^n$ where $\omega$ is the difference of the two photon virtualities
divided by their sum. However, the present data \ci{babar18} on that form factor are not accurate enough for
such an analysis \ci{KPK19}. 
Nevertheless, the values quoted in \req{eq:a2} are consistent 
with these data.

Another source of information on the gluon \da{} is provided by the inclusive $\Upsilon(1S)\to \eta' X$
decays. Ali and  Parkhomenko \ci{ali03} used the data on these decays in combination with positivity
constraints for the $\eta'g^*g$ vertex functions and found for the singlet Gegenbauer coefficients the values
\be
a_2^1(\mu_0)\=-0.04\pm 0.02\,, \qquad a_2^g(\mu_0) \= 0.12 \pm 0.05\,.
\label{eq:ali-parkhomenko}
\ee
The flavor-octet contribution was ignored in this analysis.

Wide-angle photoproduction of the $\eta'$ offers a new possibility to learn about the 2-gluon \da.
The advantage of this process over the transition form factors is that the gluon \da{} contributes
to leading order now \ci{signatures}. The analysis of this process is the subject of the present
article. For comparison we occasionally refer to $\eta$ photoproduction. Our study is timely since
the GlueX experiment at the Jefferson Lab. will measure this process.

The plan of the paper is as follows: In Sect.\ 2 we recapitulate the handbag approach to wide-angle
photoproduction of pseudoscalar mesons and $\eta$-$\eta'$ mixing. In the next section we present the
twist-2 and twist-3 subprocess amplitudes for the flavor-octet and -singlet contributions. They are
basically taken from our preceding papers \ci{KPK18,signatures,huang00,KPK21}. In Sect.\ 4 we discuss
properties and predictions for the cross section and helicity correlations for $\eta'$,
as well as $\eta$, photoproduction.
The paper ends with our summary.

\section{Handbag factorization}
The theoretical framework for wide-angle photoproduction of $\eta'$-mesons is the generalization
of the treatment of pion production~\footnote{
  The photoproduction of the $\eta$-meson has also been investigated in \ci{KPK18}. However, the
  contribution from the $gg$ Fock component has been ignored in that work which, for the $\eta$-meson,
  is a reasonable simplification.}
\ci{KPK18}. Thus, for Mandelstam variables, $s, -t$ and $-u$, much larger than $\Lambda^2$ where
$\Lambda$ is a typical hadronic scale of order $1\,\gev$, one can apply handbag factorization
in a symmetrical center-of-mass frame in which skewness, defined by
\be
\xi\=\frac{(p-p')^+}{(p+p')^+}\,,
\ee
is zero \ci{huang00,DFJK1}. The momenta of the in- and outgoing protons are denoted by $p$
and $p'$, respectively. With the help of a few plausible assumptions one can show that
the Mandelstam variables of the partonic subprocess, $\sh$, $\th$ and $\uh$, coincide
with the ones for the full process up to corrections of order $\Lambda^2/s$
\be
\th\simeq t\,, \qquad \sh\simeq s\,, \qquad \uh\simeq u\,.
\ee
The active partons, i.e.\ those which participate in the subprocess, are approximately on-shell,
move collinear with their parent hadron and carry a momentum fraction close to unity.
As in deeply virtual exclusive scattering the physical situation is that of a hard
parton-level subprocess, $\gamma q_a\to \eta'q_a$, and a soft emission and reabsorption of
quarks from the proton. Up to corrections of order $\Lambda/\sqrt{-t}$ the (light-cone)
helicity amplitudes of wide-angle photoproduction of the $\eta'$ are given by a product
of subprocess amplitudes, ${\cal H}$, and form factors which represent $1/x$-moments of
zero-skewness generalized parton distributions (GPDs):
\ba
   {\cal M}^{(i)}_{0+,\mu +}&=& \frac{e_0}{2}\,\sum_\lambda\left[ {\cal H}^i_{0\lambda, \mu\lambda}\,
                                         \Big(R^i_V(t) + 2\lambda R^i_A(t)\Big) \right.\nn\\
            &&\left. - 2\lambda \frac{\sqrt{-t}}{2m} {\cal H}^i_{0-\lambda,\mu\lambda}\,
                                         \bar{S}^i_T(t)\right]\,, \nn\\
   {\cal M}^{(i)}_{0-,\mu +} &=&\frac{e_0}{2}\,\sum_\lambda\left[\frac{\sqrt{-t}}{2m} {\cal H}^i_{0\lambda, \mu\lambda}\,
                                  R^i_T(t) \right. \nn\\
           &&\left. -2\lambda \frac{t}{2m^2}\,{\cal H}^i_{0-\lambda,\mu\lambda}\,S_S^i(t)\right]
   + e_0 {\cal H}^i_{0-,\mu +}\, S_T^i(t)\,,
\label{eq:handbag-amp}
   \ea
 where $i$ is either the flavor singlet or octet amplitude and $\mu$ denotes the helicity of
 the photon, $\lambda$ that one of the active quark and $e_0$ the positron charge. According
 to \ci{FKS1} the helicity amplitudes for $\eta'$ and $\eta$ production are given by
 \ba
    {\cal M}^{\eta'}&=& \sin{\theta_8} \,{\cal M}^{(8)} + \cos{\theta_1}\,{\cal M}^{(1)}\,, \nn\\
    {\cal M}^{\eta}&=& \cos{\theta_8} \,{\cal M}^{(8)} - \sin{\theta_1}\,{\cal M}^{(1)}\,.
 \ea
 For the mixing angles the phenomenological values \ci{FKS1}
 \be
 \theta_8\=-(21.2\pm 1.4)^\circ\,, \qquad \theta_1\=-(9.2\pm 1.4)^\circ\,,
 \label{eq:mixing-angles}
 \ee
 are adopted. These values are in reasonable agreement with the results from a recent lattice QCD
 study \ci{bali21} and from the broken hidden symmetry model \ci{jegerlehner21}. Somewhat larger differences
 to the values given in \req{eq:mixing-angles}
 have been found by Escribano and Frere \ci{escribano} in a phenomenologically study of the decays of pseudoscalar
 and vector mesons. Since their results lead to strong violations of the OZI rule we don't use their mixing
 parameters.

The form factors $R_V, R_A$ and $R_T$ 
are related to the helicity non-flip GPDs, 
$H$, $\widetilde{H}$ and $E$, at zero skewness, respectively. 
These form factors go together with quark helicity non-flip in the
subprocess, i.e.\ with the twist-2 subprocess amplitude 
${\cal H}_{0\lambda,\mu\lambda}$. 
The second set of form factors,
 $S_T$, $\bar{S}_T$ and $S_S$ are related to the helicity-flip or transversity GPDs $H_T, \bar{E}_T$ and
 $\widetilde{H}_T$, at zero skewness, respectively. 
These form factors are 
multiplied 
in \req{eq:handbag-amp} 
by the quark helicity-flip
subprocess amplitude ${\cal H}_{0-\lambda, \mu\lambda}$ 
which is of twist-3 nature~\footnote{
   Twist-3 effects can also be generated by twist-3 GPDs. However, these are expected to be small and therefore
   neglected here as in \ci{KPK18}.}.

As discussed in \ci{KPK18} the flavor-octet and singlet form factors 
$F^{(i)}_j(t)=R^{(i)}_j(t), S^{(i)}_j(t)$ 
read
($e_a$ is the charge of a flavor-a quark in units of the positron charge)
\be
F_j^{(8)} \= \frac1{\sqrt{2}} F_j^{(1)} \= \frac1{\sqrt{6}} \Big[e_u F_j^u + e_d F_j^d\Big]
\label{eq:ff}
\ee
for a proton target where the flavor form factors are
\be
F_j^a(t)\=\int_0^1 \frac{dx}{x}\,K_j^a(x,\xi=0,t)\,.
\label{eq:moment}
\ee
For charge-conjugation even mesons only valence quarks contribute~\footnote{
  In the handbag approach for wide-angle photo- and electroproduction of pseudoscalar mesons
  contribution from sea quarks are generally strongly suppressed. The sea-quark form factors drop
  typically as $F_j^{sea}(t)\sim 1/(-t)^4$ \ci{KPK18,KPK21}.}.
For a neutron target the form factors, expressed in terms of proton GPDs, $K_j^a$, read \ci{KPK18}
\be
F_{jn}^{(8)} \= \frac1{\sqrt{2}} F_{jn}^{(1)} \= \frac1{\sqrt{6}} \Big[e_u F_j^d + e_d F_j^u\Big]\,.
\label{eq:ffn}
\ee
For the numerical estimates of $\eta'$ photoproduction we take the same form factors as in \ci{KPK18}.
The $R$-type form factors are rather well-known since they are evaluated from the zero-skewness GPDs determined
in an analysis of the electromagnetic nucleon form factors \ci{DK13}. The transversity GPDs, $H_T$ and $\bar{E}_T$,
are extracted from data on deeply-virtual pion electroproduction at low $-t$ \ci{GK6}. Their large $-t$
behavior is adjusted to the CLAS data on wide-angle $\pi^0$ photoproduction \ci{clas17}. The form factor
$S_S$ is assumed to be $\bar{S}_T/2$. The $S$-type form factors demand improvements. However, these form
factors are not implausible as a comparison with preliminary GlueX data \ci{kamal21} on $\eta$ photoproduction
reveals (at $s=16.36\,\gev^2$), see Sect.\ \ref{sec:results}.

\section{The subprocess amplitudes}
\begin{figure}
      \centering
      \includegraphics[width=0.6\tw,bb=70 493 536 720,clip=true]{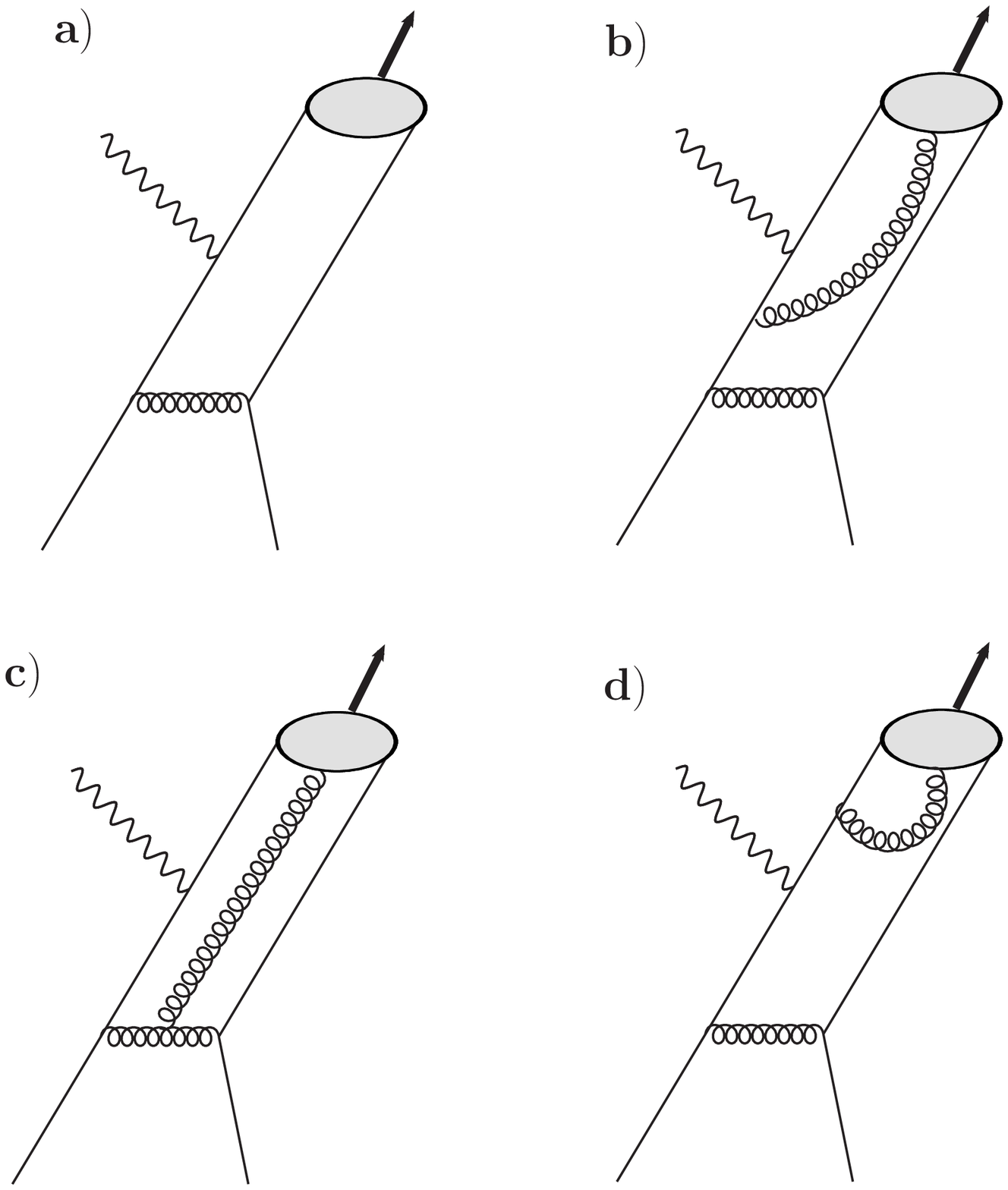}
      \hspace*{0.03\tw}
      \includegraphics[width=0.24\tw]{gg-graphs.epsi}
      \caption{Typical Feyman graphs for wide-angle photoproduction of the $\eta' (\eta)$ meson: a) twist-2
        and twist-3 for $\gamma q\to (q\bar{q}) q$, b) twist-3 for $\gamma q\to (q\bar{q}g) q$,
        c) twist-2 for $\gamma q\to (gg) q$. }
      \label{fig:graphs}
 \end{figure}
Typical leading-order Feynman graphs for the subprocess $\gamma q_a\to \eta' q_a$ are shown
in Fig.\ \ref{fig:graphs}. We stress that in the soft meson and nucleon matrix elements defining
the \da s and GPDs, we are using light-cone gauge. 
The twist-2 subprocess amplitudes read~\footnote{
  We remark that in deeply-virtual electroproduction of the $\eta'$ meson the gluon-gluon contribution
  to the subprocess amplitudes is suppressed by $\th/Q^2$ in the generalized Bjorken regime, see \ci{KPK03}.}  
\ci{signatures,huang00}
\ba
{\cal H}^{(8),tw2}_{0\lambda,\mu\lambda}&=&\sqrt{2}\pi\als(\mu_R) f_8\, \frac{C_F}{N_C}\, \frac{\sqrt{-\th}}{\sh\uh}
   \,\langle 1/\tau\rangle_{q8}\,\Big[(1+2\lambda\mu)\sh - (1-2\lambda\mu)\uh\Big]\,,\nn\\
{\cal H}^{(1),tw2}_{0\lambda,\mu\lambda}&=&\sqrt{2}\pi\als(\mu_R) f_1\, \frac{C_F}{N_C}\, \frac{\sqrt{-\th}}{\sh\uh}
                          \,\Big[\langle 1/\tau\rangle_{q1}-\langle 1/\tau^2\rangle_g\Big] \nn\\
                        &&\hspace*{0.25\tw} \times  \Big[(1+2\lambda\mu)\sh - (1-2\lambda\mu)\uh\Big]\,.   
\label{eq:twist2-subamp}
\ea
where
\ba
\langle 1/\tau \rangle_{qi}&=& \int_0^1 \frac{d\tau}{\tau}\,\Phi_{qi}(\tau,\mu_F)\simeq 3\,\big[1+a_2^i(\mu_F)\big]\,,
                                                                                           \nn\\
\langle 1/\tau^2\rangle_g &=& \int_0^1 \frac{d\tau}{\tau^2}\,\Phi_g(\tau,\mu_F) \simeq - 25\,a_2^g(\mu_F)
\label{eq:moments}
\ea
for the truncated Gegenbauer expansions of the \da s defined in \req{eq:das}. As usual $C_F=(N_C^2-1)/(2N_C)$ is a
color factor, $N_C$ denotes the number of colors and $\mu_R$ is the renormalization scale.

The octet and singlet decay constants are taken from \ci{FKS1} ($f_\pi=132\,\mev$)~\footnote{
  The mixing angles and the decay constants can solely be expressed in terms of particle masses.
  These expressions lead to values for these parameters in very good agreement to those given in
  \req{eq:mixing-angles} and \req{eq:decay-constants}, see \ci{kroll}.}:
\be
f_8\= (1.26\pm 0.06)\, f_\pi\,, \qquad f_1\= (1.17\pm 0.04)\,f_\pi\,.
\label{eq:decay-constants}
\ee
In contrast to the mixing angles and $f_8$, the singlet decay constant is factorization
scale dependent but this is an NLO effect \ci{espriu}  which we ignore for consistency as has been done in \ci{FKS1}.
In a recent lattice QCD study \ci{bali21} the scale dependence of $f_1$ has indeed been observed
and its value at the scale $1\,\gev$ agrees very well with the one quoted in \req{eq:decay-constants}.
Thus, one may assume that the latter value holds at $\mu_F\simeq 1\,\gev$. 
We would like to add that also $f_8$ determined in \ci{bali21} is in good agreement with the above given value.

As shown in \ci{KPK18} and \ci{KPK21} the twist-3 contribution plays 
an important role in the photoproduction of pseudoscalar mesons. 
The flavor-octet case, with the \da{} $\Phi_{38}$, is discussed in detail in \ci{KPK18}. 
The flavor-singlet component, to which a \da{} $\Phi_{31}$ contributes,
has a similar structure. 
The absence of 2- and 3-gluon twist-3 contributions
can be understood as follows: Consider an $\eta'$-meson moving rapidly along the 3-axis. 
A light-cone wave function of its $n$-parton Fock component 
with orbital angular momentum projection onto the 3-direction, $l_3$, 
has the dimension $[\mbox{mass}]^{(n+|l_3|-1)}$ \ci{ji03}.
 In a hard exclusive process this dimension has to be balanced by corresponding inverse powers of the hard scale,
 $\sqrt{s}$ in our case. The associated \da , i.e. the light-cone \wf{} integrated upon the parton transverse momenta,
 is of twist $n+|l_3|$ nature. Now, consider an $\eta'$ Fock component consisting of two massless gluons. 
Its total helicity, 
i.e.\ its total spin projection on the 3-direction, $s_3$, 
is either zero or $\pm 2$. 
Hence, for a spin-0 hadron
 either $|l_3|=0$ or
 2 is required. The $l_3=0$ case leads to the twist-2 \da{} given in \req{eq:das}, which contributes to the subprocess
 amplitude ${\cal H}^{(1),tw2}$  quoted in \req{eq:twist2-subamp}. The $|l_3|=2$ one is of twist-4 
 nature and is neglected here in this work. For a 3-gluon state one has either $s_3=\pm 1$ or 3, demanding $|l_3|=1$
 or 3, respectively. The associated \da s are of twist-4 or higher nature. Hence, there is neither a 2-gluon nor
 a 3-gluon twist-3 \da{} for the $\eta'$ meson. These observations are in accordance with the glueball spectrum
 \ci{fritzsch,boulanger}.

 The 3-body flavor-singlet \da{}, $\Phi_{31}$, is completely unknown as yet. The situation for the
flavor-octet \da{} is somewhat better. Flavor symmetry tells us that $\Phi_{38}$ should be close to $\Phi_{3\pi}$
which is supported by a QCD sum rule study \ci{ball98}. Thus, the best one can do at present is to assume
\be
\Phi_{38}(\tau_a,\tau_b,\tau_g)\=\Phi_{31}(\tau_a,\tau_b,\tau_g)\simeq \Phi_{3\pi}(\tau_a,\tau_b,\tau_g)\,.
\label{eq:twist3-das}
\ee
We stress that the assumption on $\Phi_{31}$ is a pure guess. The twist-3 pion \da{} is taken from \ci{KPK18} 
where a truncated Jacobi-polynomial expansion \ci{braun} has been employed. Therefore, we have
\ba
\Phi_{38}(\tau_a,\tau_b,\tau_g,\mu_F)&=& \Phi_{31}(\tau_a,\tau_b,\tau_g,\mu_F)   \nn\\
                &=&  360 \tau_a\tau_b\tau_g^2\Big[ 1 + \omega_{1,0}(\mu_F)\frac12(7\tau_g-3) \nn\\
        && + \,\omega_{2,0}(\mu_F) (2-4\tau_a\tau_b-8\tau_g+8\tau_g^2)  \nn\\
 && +\, \omega_{1,1}(\mu_F)(3\tau_a\tau_b-2\tau_g+3\tau_g^2)\Big]\,.
\ea
The variable $\tau_g$ refers to the fraction of the meson momentum the gluon carries.
The expansion coefficients are \ci{KPK18}:
\be
\omega_{1,0}(\mu_0)\=-2.55\,, \quad  \omega_{1,1}(\mu_0)\=0\,. 
\label{eq:omega}
\ee
As in \ci{KPK18,KPK21} we consider the coefficient $\omega_{2,0}$ as a free parameter fitted to available data, in the
present case to the preliminary GlueX data \ci{kamal21} on $\eta$ photoproduction~\footnote{
  For comparison we repeat the value of this coefficient for the pion: $\omega_{20}=8.0$ if evolution of \da s is taken into
  account and 10.3 for the fixed-scale calculation.}.
The expansion coefficients $\omega_{2,0}$ and $\omega_{1,1}$ mix under evolution.
The 3-body twist-3 \da s are to be multiplied by the normalizations $f_{3i}$ ($i=8,1$), defined such that the
corresponding \da s integrated upon the momentum fractions are unity. 
For the normalizations we take   
similarly to \ci{KPK18}   
 \be
 f_{38}(\mu_0)\=0.86 f_{3\pi}(\mu_0)\,, \qquad f_{31}(\mu_0)\=0.86 f_{3\pi}(\mu_0)\,,
\label{eq:twist3-decay-constants}
 \ee
 with
\be
f_{3\pi}(\mu_0)\=0.004 \pm 0.001\, \gev^2\,.
\label{eq:f3pi}
\ee
The normalizations $f_{38}$ and $f_{3\pi}$ are supported by QCD sum rule studies \ci{ball98,braun}.
The value of $f_{31}$ is a supposition which is, to some extent, justified by the fair
agreement of our predictions with the preliminary GlueX data \ci{kamal21} on wide-angle photoproduction
of the $\eta$ meson. 

There are also two 2-body twist-3 \da s, $\Phi_{pi}$ and $\Phi_{\sigma i}$. They are not needed explicitly here,
since, due to the equation of motion \ci{KPK18}, the complete twist-3 subprocess amplitude can solely be
expressed by the 3-body \da{}~\footnote{
  This result implies that the Wandzura-Wilczek approximation is zero in hard wide-angle
  photoproduction of pseudoscalar mesons.}. It reads 
\ba
   {\cal H}^{(i),tw3}_{0-\lambda,\mu\lambda}&=& 2\sqrt{2}\pi\,(2\lambda - \mu)\,\als(\mu_R)\frac{C_F}{N_C}\,f_{3i}(\mu_F)\,
                                       \frac{\sqrt{-\uh\sh}}{\sh^2\uh^2}\,  \int_0^1 d\tau\,\nn\\
             &\times& \int_0^{\taub} \frac{d\tau_g}{\tau_g} \Phi_{3i}(\tau,\taub-\tau_g,\tau_g,\mu_F)
          \left[\left(\frac1{\taub^2}-\frac1{\taub(\taub-\tau_g)}\right)\Big(\sh^2+\uh^2\Big) \right.\nn\\
           &&\left. - \left(1-\frac12\frac{N_C}{C_F}\right)\left(\frac1{\tau}+\frac1{\taub-\tau_g}\right)\,
                             \frac{\th(\sh+\uh)}{\tau_g}\right]
\label{eq:twist3-subamp}
\ea                          
The subprocess amplitudes, ${\cal H}^{(i),tw2}$ 
(\ref{eq:twist2-subamp})
and ${\cal H}^{(i),tw3}$, satisfy current conservation and
are gauge invariant in QCD.

\section{Predictions on and properties of $\eta'$ photoproduction}
\label{sec:results}

Before we present numerical results on $\eta'$ photoproduction an important issue, the energy
dependence of the cross section, is to be discussed. According to leading-twist dimensional counting
the cross sections for photoproduction of pseudoscalar mesons should scale as $s^{-7}$ at fixed
$\cos{\theta}$ where $\theta$ is the scattering angle in the center-of-mass system.  Further energy
dependence comes from 
the running of $\alpha_S$,
the evolution of the decay constants and the \da s
as well as from the twist-3 contribution which is suppressed by $1/\sqrt{s}$ at the amplitude level
compared to the twist-2 one, cf.\ \req{eq:twist2-subamp} and \req{eq:twist3-subamp}. The soft form
factors also contribute to the extra energy-dependence except they, including possible prefactors of
$\sqrt{-t}$ and $t$ appearing in \req{eq:handbag-amp}, fall $\propto 1/t^2$. For the present parameterization
\ci{KPK18,KPK21} the form factors fall slightly faster and the $d$-quark form factors even faster than the
$u$-quark ones. In the range of $s$ between, say, $10$ and $20\,\gev^2$ our cross sections for $\eta'$
and $\eta$ photoproduction effectively fall about as $s^{-9}$. This is perhaps too strong.
We stress that $\eta'$ photoproduction has not been measured yet in the wide-angle region at
  high energies and for $\eta$ photoproduction we only have at disposal the preliminary GlueX data \ci{kamal21}
  at a single energy $s=16.36\,\gev^2$. Only for pion photoproduction there are data for several values
  of large $s$ available from an old SLAC experiment \ci{slac} which, for $\pi^+$ production, are in agreement
  with the dimensional counting result of a $s^{-7}$ drop. One may however wonder why the QCD logarithms from the
  evolution and from the running of $\als$ are not perceptible. For $\pi^0$ production the situation is unclear
  since the SLAC data are not compatible with the recent CLAS measurement \ci{clas17}.
  
\begin{figure}
      \centering
      \includegraphics[width=0.47\tw]{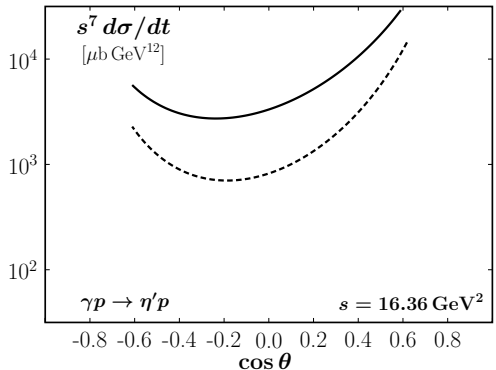}
      \caption{The $\eta'$ cross section, scaled by $s^7$, at $s=16.36\,\gev^2$ with (dashed line) and
        without evolution (solid line). The twist-3 expansion coefficient $\omega_{2,0}$ is $6.0$.}
     \label{fig:evolution}
  \end{figure}

  In this 
situation we follow the remedy advocated for in \ci{KPK21} and evaluate the cross sections at
  the fixed scale $\mu_R=\mu_F=1\,\gev$. In this case the effective energy dependence of the cross sections is milder,
  about $s^{-8}$. This can be seen from Fig.\ \ref{fig:evolution} where we display the $\eta'$ cross section
  evaluated with the fixed  and with the running scale at $s=16.36\,\gev^2$. This value of $s$ is chosen in order
  to facilitate the comparison with the GlueX data on $\eta$ photoproduction \ci{kamal21} as soon as they are published.
  For the evaluation of the cross sections we use the mixing angles \req{eq:mixing-angles}, the twist-2 parameters
  \req{eq:a2}  and \req{eq:decay-constants} as well as the twist-3 ones, \req{eq:omega} and \req{eq:twist3-decay-constants}.
  As in our previous work \ci{KPK18,KPK21} for the running scale 
we take $\mu_F=\mu_R=\th \uh/\sh$ and evaluate $\als(\mu_R)$ from the one-loop
  approximation with $\LQCD=0.22\,\gev$ and $n_f=4$ flavors. The anomalous dimensions required for the evolution of the
  various \da s have been derived in \ci{grozin,ball98,braun} and are systematized in \cite{KPK13,KPK21}.
  We see that without evolution the cross section is substantially larger than with evolution.
  All results shown in the following are evaluated at the fixed scale.
  As soon as sufficient data on these cross sections will become available the issue of the scale dependence is
  to be taken up again.

\begin{figure}
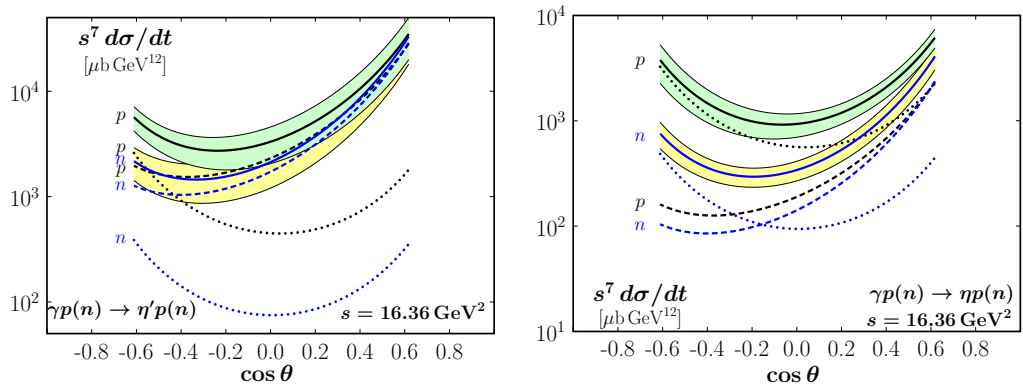

      \centering
      \includegraphics[width=0.47\tw]{fig-dsdt-etap-tw23.epsi}
      \hspace*{0.02\tw}
      \includegraphics[width=0.47\tw]{fig-dsdt-eta-tw23.epsi}
      \caption{$\eta'$ (left) and $\eta$ (right) cross sections, scaled by $s^7$, at $s=16.36\,\gev^2$ and
        $\omega_{2,0}=6.0$. Solid (dashed, dotted) lines are for the full (twist 2, twist 3) cross sections 
        for a proton (labeled p) and  neutron (labeled n) target. The parameters are \req{eq:a2}, \req{eq:omega} and
        \req{eq:omega20}. The shaded bands represent the parametric errors of the full cross sections.}
      \label{fig:tw23}
  \end{figure}

  In Fig.\ \ref{fig:tw23} we display the $\eta'$ cross sections for a proton and a neutron target
  at $s=16.36\,\gev^2$. For comparison we also show the analogous $\eta$ cross section.
  The expansion coefficient $\omega_{2,0}$ of the twist-3 \da s $\Phi_{38}=\Phi_{31}$, fitted to the 
  available $\gamma p\to \eta p$ cross section data~\footnote{
      The GlueX collaboration did not give us the permission to show their data.}
    \ci{kamal21}, is
\be
\omega_{2,0}(\mu_0)\=6.0\,.
\label{eq:omega20}
\ee
  We also show in Fig.\ \ref{fig:tw23} the pure twist-2 and twist-3 contributions separately. 
As is evident from this plot twist 2 predominates 
$\eta'$ photoproduction for $\cos{\theta}\geq 0$ 
both for proton and neutron target. 
In the backward region
  the twist-2/twist-3 interference is substantial and, for a proton target, twist 3 is large for
  $\cos{\theta}\lsim -0.4$. 
On the other hand, twist 3 predominates $\eta$ photoproduction 
off protons except in
 the very forward region where the twist-2 contribution 
is of about the same size as the twist-3 one. For a
  neutron target the forward hemisphere is dominated by twist 2, the backward one by twist 3. The twist-2/twist-3
  interference is strong in both the regions.
In Fig.\ \ref{fig:tw23} we also display error bands. They represent the parametric errors
of the cross sections evaluated from all errors mentioned in the text as well as from those of the soft
form factors \ci{KPK18,DK13}. For $\eta'$ production the most important error is that of the twist-2 parameter
$a_2^g$ (see Eq.\ \req{eq:a2}) except for $\cos{\theta}\lsim -0.4$ where also the error of $f_{3\pi}$ (see Eqs.\
\req{eq:twist3-decay-constants}, \req{eq:f3pi}) matters. The latter error influences strongly the error bands
for $\eta$ production for all relevant scattering angles.

\begin{figure}
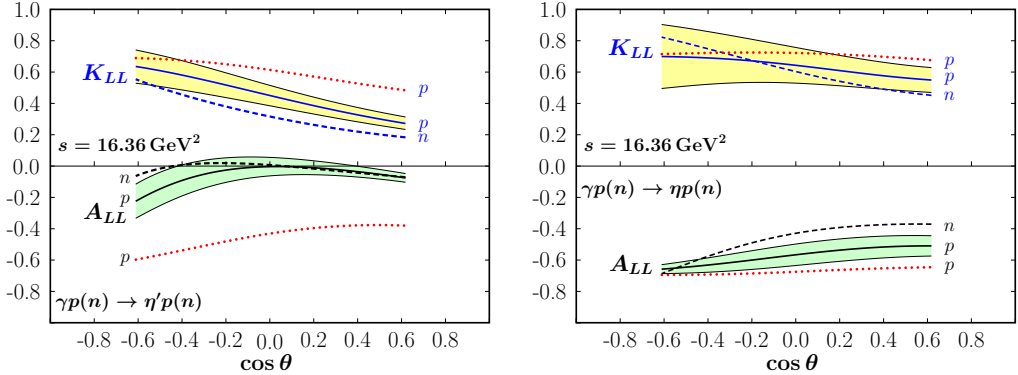

      \centering
      \includegraphics[width=0.47\tw]{fig-ALL-KLL-etap.epsi}
      \hspace*{0.02\tw}
      \includegraphics[width=0.47\tw]{fig-ALL-KLL-eta.epsi}
      \caption{The helicity correlations $A_{LL}$ and $K_{LL}$ for $\eta'$ (left) and $\eta$ (right) photoproduction
        off protons (solid lines) and off neutrons (dashed lines) at $s=16.36\,\gev^2$. Solid lines: using the parameters
        \req{eq:a2}, \req{eq:omega} and \req{eq:omega20}. Dotted lines: using the twist-2
        singlet coefficients \req{eq:ali-parkhomenko} and $\omega_{20}(\mu_0)=7.3$. The shaded bands represent
      the parametric errors of $A_{LL}$ and $K_{LL}$ for $\eta$ and $\eta'$ photoproduction off protons.}
      \label{fig:ALL-KLL}
   \end{figure}

  Interesting spin-dependent observables are the correlations between the helicities of the photon and that of
  either the incoming or the outgoing nucleon, $A_{LL}$ and $K_{LL}$, respectively. As we showed in
  \ci{KPK18}, for twist 2, one has
  \be
  A_{LL}^{tw2}\= K_{LL}^{tw2}
  \ee
  whereas for twist 3
  \be
  A_{LL}^{tw3}\= -K_{LL}^{tw3}
  \ee
  holds. In Fig.\ \ref{fig:ALL-KLL} we display these correlations for $\eta$ and $\eta'$ photoproduction~\footnote{
    The ${\cal M}_{0\lambda',\mu\lambda}$ are light-cone helicity amplitudes. For comparison with
    experimental data on spin-dependent observables the use of the ordinary helicity basis is more convenient.
    The transform of the light-cone helicity amplitudes to the ordinary helicity ones is discussed for
    photoproduction of pseudoscalar mesons in \ci{KPK18}.}.
  The pattern of curves is very different for the two cases. For $\eta$ photoproduction
  approximate mirror symmetry is to be seen, implying strong twist-3 contributions in accordance with the
  behavior of the corresponding cross section, see Fig.\ \ref{fig:tw23}.  For $\eta'$ photoproduction, on the
  other hand, the correlation $A_{LL}$ is much smaller in absolute value than $K_{LL}$. This indicates a
  larger significance of  twist-2. 
  We emphasize that the twist-2/twist-3 interference is more important
  for the helicity correlations than for the cross sections.

  The results for cross sections and helicity correlations at $s=16.36\,\gev^2$ are characteristic for
  the energy range $10\,\gev^2 \lsim s \lsim 20\,\gev^2$ in the wide-angle region defined by $-t$ and $-u$
  larger than about $2.5\,\gev^2$ and  $|\cos{\theta}|\lsim 0.6$. The relative order of twist-2 and twist-3
  contributions remains about the same in the wide-angle region although for increasing $s$ twist-2
  becomes more important. Asymptotically the twist-2 contribution dominates.

   It remains to examine the influence of the unknown flavor-singlet twist-3 \da{} for which we made the
   assumptions \req{eq:twist3-das} and \req{eq:twist3-decay-constants}. In order to check that we drastically
   enlarge the flavor-singlet twist-3 contribution by multiplying it by $1.3$. Doing so we observe that,
   for $\cos{\theta}\simeq 0.6$ the $\eta'$ cross section off protons (neutrons) changes by about $\pm 10 (7)\%$.
   The effect on the cross section increases with decreasing $\cos{\theta}$. It amounts to about $\pm 20\, (12)\%$
   at $\cos{\theta}\simeq 0$ and to about $\pm 50\, (30)\%$ at $\cos{\theta}\simeq -0.6$.

   \begin{figure}
      \centering
      \includegraphics[width=0.47\tw]{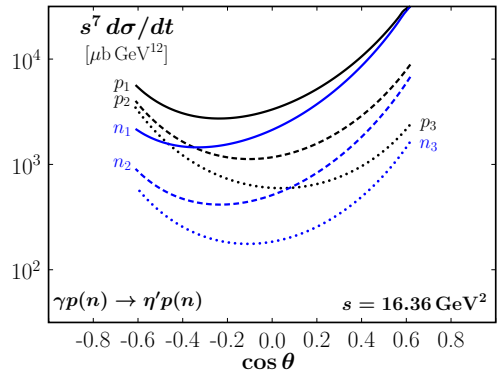}
      \hspace*{0.02\tw}
      \caption{The cross section for $\eta'$ photoproduction off protons and off neutrons at $s=16.36\,\gev^2$ for 
        three different scenarios. Solid lines: expansion coefficients according to \req{eq:a2}, \req{eq:omega}
        and \req{eq:omega20}; dashed lines: using the twist-2 singlet coefficients \req{eq:ali-parkhomenko} and
        $\omega_{20}(\mu_0)=7.3$; dotted lines: using $a_2^8(\mu_0)=a_2^1(\mu_0)=a_{2\pi}(\mu_0)$, $a_2^g(\mu_0)=0$ and
        $\omega_{20}(\mu_0)=7.8$.}
      \label{fig:etap-scen}
   \end{figure}

  The above discussion in combination with Figs.\ \ref{fig:tw23} and \ref{fig:ALL-KLL} makes it clear that
  photoproduction of the $\eta'$ meson in the forward hemisphere seems to be suitable for a determination of
  the twist-2 gluon-gluon \da. In order to see whether this result also holds for flavor-singlet \da s
  that are not close to the one given in Eq.\ \req{eq:a2}  we next evaluate the $\eta'$ and $\eta$ cross sections
  for the twist-2 flavor-singlet expansion coefficients \req{eq:ali-parkhomenko} but keeping the value $a_2^8(\mu_0)$
  given in \req{eq:a2}. All other parameters remain unchanged except of $\omega_{2,0}$ which is taken to be
   7.3 in order to have still fair agreement with the GlueX data \ci{kamal21} on the $\eta$ cross
   section. The results on the $\eta$ cross section for this set of parameters lie within the error band shown in
   Fig.\ \ref{fig:tw23}. The predictions on the $\eta'$ cross section are shown in Fig.\ \ref{fig:etap-scen}. We see
   that for this second scenario the $\eta'$ cross section is substantially smaller then that one obtained
   from the flavor-singlet Gegenbauer coefficients given in Eq.\ \req{eq:a2}. Nevertheless we have twist-2 dominance
   in the forward hemisphere and the helicity correlations differ only mildly from those obtained with the first set
   of parameters, see Fig.\ \ref{fig:ALL-KLL}. At present we cannot say which of the scenarios is to be favored.

   For comparison we also display in Fig.\ \ref{fig:etap-scen}
   results of a third scenario for which the gluon \da{} is assumed to be zero at the initial scale $\mu_0$
   and $a_2^8(\mu_0)=a_2^1(\mu_0)=a_{2\pi}(\mu_0)$. For the second Gegenbauer coefficient of the pion \da{}
   we take a recent lattice QCD result \ci{braun15}: $a_{2\pi}(\mu_0)=0.1364$. We now take $\omega_{2,0}=7.8$ 
   which, as for the other scenarios, also leads to fair agreement with the GlueX data \ci{kamal21}.
   For this extreme scenario the $\eta'$ cross section is even smaller than for the second one. The twist-3
   contribution to the $\eta'$ cross section now also dominates in the forward hemisphere. In accordance with that
   the helicity correlations are now  similar to those of $\eta$ production.

\section{Summary}
We have investigated wide-angle photoproduction of $\eta'$ mesons at high energies within the handbag
approach in which the process amplitudes factorize into hard perturbatively calculable subprocesses and
soft form factors representing $1/x$-moments of GPDs. The soft formfactors for given flavors are taken
from our analysis of pion photoproduction \ci{KPK18}. For the evaluation of the subprocess amplitudes
the twist-2 and twist-3 \da s for flavor-singlet and -octet components of the $\eta'$ mesons are needed.
Whereas fair knowledge of the flavor-octet \da s is available 
the twist-2 flavor-singlet \da s,
the quark-antiquark one as well as the gluon-gluon one, 
are poorly known. The available information mainly
comes from a NLO analysis of the $\eta$- and $\eta'$-photon transition form factor. The twist-3
flavor-singlet \da{} is yet totally unknown. Assuming $\Phi_{38}=\Phi_{31}$ and $f_{38}=f_{31}$, we have found
that $\eta'$ photoproduction is dominated by the twist-2 contribution in the forward hemisphere. Thus,
the twist-3 flavor-singlet \da{} plays only a minor role for $\eta'$ photoproduction in that region. This
is to be contrasted with $\eta$ photoproduction where the twist-3 contributions play the leading role. We
have found that, with our assumption on $\Phi_{31}$ and $f_{31}$, reasonable agreement with the preliminary
GlueX data \ci{kamal21} on $\eta$ photoproduction at $s=16.36\,\gev^2$ is obtained. 
We have shown that the cross sections for $\eta'$ photoproduction off protons or neutrons are very sensitive
to the twist-2 gluon \da{} in particular for $\cos{\theta}\geq 0$. We emphasize
that, in contrast to the meson-photon transition form factors, the contribution from the gluon-gluon Fock
component is not suppressed by $\als$. It will be interesting to confront our predictions with
the forthcoming data from the Jefferson Lab GlueX experiment and to see what we can learn on the twist-2
flavor-singlet \da s. The planned measurement of the helicity correlation $A_{LL}$ for the processes of
interest by the Jefferson Lab Frozen Spin experiment will provide additional information on these \da s.
With sufficient data on wide-angle $\eta'$ (and $\eta$) photoproduction at disposal the issue of the scale
dependence is to be resumed.

{\it Acknowledgment} We thank Igor Strakovsky for informing us about the GlueX measurement of the $\eta$
photoproduction cross section.
This publication is supported
by the Croatian Science Foundation project IP-2019-04-9709,
and by the EU Horizon 2020 research and innovation programme, STRONG-2020
project, under grant agreement No 824093.
   


\begin{thebibliography}{99}

\bibitem{KPK18} P.~Kroll and K.~Passek-Kumeri\v{c}ki,
Phys. Rev. D \textbf{97} (2018) no.7, 074023
[arXiv:1802.06597 [hep-ph]].

\bibitem{GK6} S.~V.~Goloskokov and P.~Kroll,
Eur. Phys. J. A \textbf{47} (2011), 112
[arXiv:1106.4897 [hep-ph]].

\bibitem{bedlinskiy} I.~Bedlinskiy \textit{et al.} [CLAS],
Phys. Rev. Lett. \textbf{109}, 112001 (2012)
[arXiv:1206.6355 [hep-ex]].

\bibitem{defurne} M.~Defurne \textit{et al.} [Jefferson Lab Hall A],
Phys. Rev. Lett. \textbf{117}, no.26, 262001 (2016)
[arXiv:1608.01003 [hep-ex]].

\bibitem{ali03} A.~Ali and A.~Y.~Parkhomenko,
Eur. Phys. J. C \textbf{30}, 183-195 (2003)
[arXiv:hep-ph/0304278 [hep-ph]].

\bibitem{cisek21} A.~Cisek and A.~Szcurek,
Phys. Rev. D \textbf{103} (2021), 114008
[arXiv:2103.08954 [hep-ph]].

\bibitem{harland11}L.~A.~Harland-Lang, V.~A.~Khoze, M.~G.~Ryskin and W.~J.~Stirling,
Eur. Phys. J. C \textbf{71}, 1714 (2011)
[arXiv:1105.1626 [hep-ph]].

\bibitem{ball07} P.~Ball and G.~W.~Jones,
JHEP \textbf{08}, 025 (2007)
[arXiv:0706.3628 [hep-ph]].

\bibitem{bass18} S.~D.~Bass and P.~Moskal,
Rev. Mod. Phys. \textbf{91}, no.1, 015003 (2019)
[arXiv:1810.12290 [hep-ph]].   

\bibitem{KPK03} P.~Kroll and K.~Passek-Kumeri\v{c}ki,
Phys. Rev. D \textbf{67} (2003), 054017
[arXiv:hep-ph/0210045 [hep-ph]].

\bibitem{KPK13} P.~Kroll and K.~Passek-Kumeri\v{c}ki,
J. Phys. G \textbf{40} (2013), 075005
[arXiv:1206.4870 [hep-ph]]. 

\bibitem{CLEO} J.~Gronberg \textit{et al.} [CLEO],
Phys. Rev. D \textbf{57}, 33-54 (1998)
[arXiv:hep-ex/9707031 [hep-ex]].

\bibitem{L3} M.~Acciarri \textit{et al.} [L3],
Phys. Lett. B \textbf{418}, 399-410 (1998)

\bibitem{Babar} P.~del Amo Sanchez \textit{et al.} [BaBar],
Phys. Rev. D \textbf{84}, 052001 (2011)
[arXiv:1101.1142 [hep-ex]].

\bibitem{FKS1} T.~Feldmann, P.~Kroll and B.~Stech,
Phys. Rev. D \textbf{58} (1998), 114006
[arXiv:hep-ph/9802409 [hep-ph]].

\bibitem{KPK19} P.~Kroll and K.~Passek-Kumeri\v{c}ki,
Phys. Lett. B \textbf{793}, 195-199 (2019)
[arXiv:1903.06650 [hep-ph]].

\bibitem{grozin} V.~N.~Baier and A.~G.~Grozin,
Nucl. Phys. B \textbf{192}, 476-488 (1981).

\bibitem{babar18}  J.~P.~Lees \textit{et al.} [BaBar],
Phys. Rev. D \textbf{98}, no.11, 112002 (2018)
[arXiv:1808.08038 [hep-ex]].


\bibitem{signatures} H.~W.~Huang, R.~Jakob, P.~Kroll and K.~Passek-Kumeri\v{c}ki,
  Eur.\ Phys.\ J.\ C {\bf 33}, 91 (2004)
  [hep-ph/0309071].

\bibitem{huang00} H.~W.~Huang and P.~Kroll,
Eur. Phys. J. C \textbf{17}, 423-435 (2000)
[arXiv:hep-ph/0005318 [hep-ph]].

\bibitem{KPK21} P.~Kroll and K.~Passek-Kumeri\v{c}ki, 
Phys. Rev. D \textbf{104} (2021) no.5, 054040
[arXiv:2107.04544 [hep-ph]].

\bibitem{DFJK1} M.~Diehl, T.~Feldmann, R.~Jakob and P.~Kroll,
Eur. Phys. J. C \textbf{8}, 409-434 (1999)
[arXiv:hep-ph/9811253 [hep-ph]].

\bibitem{bali21} G.~S.~Bali, V.~Braun, S.~Collins, A.~Sch\"afer and J.~Simeth,
JHEP \textbf{08}, 137 (2021)
[arXiv:2106.05398 [hep-lat]].
  

\bibitem{jegerlehner21} M.~Benayoun, L.~DelBuono and F.~Jegerlehner,
[arXiv:2105.13018 [hep-ph]]. 

\bibitem{escribano}  R.~Escribano and J.~M.~Frere,
JHEP \textbf{06}, 029 (2005)
[arXiv:hep-ph/0501072 [hep-ph]].

  
\bibitem{DK13} M.~Diehl and P.~Kroll,
  Eur.\ Phys.\ J.\ C {\textbf 73}, no. 4, 2397 (2013)
  [arXiv:1302.4604 [hep-ph]].

\bibitem{clas17} M.~C.~Kunkel \textit{et al.} [CLAS],
Phys. Rev. C \textbf{98}, no.1, 015207 (2018)
[arXiv:1712.10314 [hep-ex]].

\bibitem{kamal21} M.~Kamal, talk presented at the APS GDH workshop, April 2021.

\bibitem{kroll} P.~Kroll,
Mod. Phys. Lett. A \textbf{20}, 2667-2684 (2005)
[arXiv:hep-ph/0509031 [hep-ph]].

\bibitem{espriu} D.~Espriu and R.~Tarrach,
Z. Phys. C \textbf{16}, 77 (1982).  

\bibitem{ji03} X.~d.~Ji, J.~P.~Ma and F.~Yuan,
Phys. Rev. Lett. \textbf{90}, 241601 (2003)
[arXiv:hep-ph/0301141 [hep-ph]].

\bibitem{fritzsch} H.~Fritzsch and P.~Minkowski,
Nuovo Cim. A \textbf{30}, 393 (1975)

\bibitem{boulanger} N.~Boulanger, F.~Buisseret, V.~Mathieu and C.~Semay,
Eur. Phys. J. A \textbf{38}, 317-330 (2008)
[arXiv:0806.3174 [hep-ph]].


\bibitem{ball98} P.~Ball,
JHEP \textbf{01} (1999), 010
[arXiv:hep-ph/9812375 [hep-ph]].

\bibitem{braun} V.~M.~Braun and I.~E.~Filyanov,
  Z. Phys. C \textbf{48}, 239-248 (1990).

\bibitem{slac} L.~Anderson, D.~Gustavson, D.~Ritson, G.~A.~Weitsch, H.~J.~Halpern, R.~Prepost,
             D.~H.~Tompkins and D.~E.~Wiser,
Phys. Rev. D \textbf{14}, 679 (1976).

\bibitem{braun15} V.~M.~Braun, S.~Collins, M.~G\"ockeler, P.~P\'erez-Rubio, A.~Sch\"afer, R.~W.~Schiel
  and A.~Sternbeck,
Phys. Rev. D \textbf{92}, no.1, 014504 (2015)
[arXiv:1503.03656 [hep-lat]].


\end{thebibliography}
\end{document}